\documentclass[a4paper,11pt]{article}
\usepackage{pos}
\usepackage{gensymb}
\usepackage{wrapfig}
\usepackage{lineno}
\def\dg{\hbox{$^\circ$}}

\def\utw{\smash{\rlap{\lower5pt\hbox{$\sim$}}}}
\def\udtw{\smash{\rlap{\lower6pt\hbox{$\approx$}}}}

\title{Correction method applied to MC simulated LST images affected by clouds}

\author*[a,b]{Natalia Żywucka}
\author[a,\ddag]{Julian Sitarek}
\author[a]{Dorota Sobczyńska}
\author[c,d]{Mario Pecimotika}
\author[e]{Dario Hrupec}
\author[c]{Dijana Dominis Prester}
\author[c]{Lovro Pavletić}
\author[c]{Saša Mićanović}

\affiliation[a]{Department of Astrophysics, University of Łódź, Pomorska 149, Łódź, Poland }

\affiliation[b]{North-West University, Potchefstroom, South Africa}

\affiliation[c]{Faculty of Physics, University of Rijeka, Radmile Matejčić 2, Rijeka, Croatia}

\affiliation[d]{Institute Ruđer Bošković, Bijenička cesta 54, Zagreb, Croatia}

\affiliation[e]{Department of Physics, J. J. Strossmayer University of Osijek, Trg Ljudevita Gaja 6, Osijek, Croatia}

\affiliation[\ddag]{Now at Institute for Cosmic Ray Research (ICRR), The University of Tokyo, Kashiwa, 277-8582 Chiba}

\emailAdd{natalia.zywucka@fis.uni.lodz.pl}
\abstract{We present the results of a preliminary study of a correction method applied to the Imaging Atmospheric Cherenkov Telescope images affected by clouds. The studied data are Monte Carlo simulations made with CORSIKA, imitating the very high energy events registered by the Large-Sized Telescopes, a type of telescope within the future Cherenkov Telescope Array. We implement the cloud correction method in the \texttt{ctapipe}/\texttt{lstchain} analysis framework. The correction is based on a simple geometrical model of the emission. We show the effect of the correction method on the image parameters and the stereo-reconstructed shower parameters.}

\FullConference{  
High Energy Astrophysics in Southern Africa 2022 - HEASA2022\\
  28 September - 1 October 2022\\
  Brandfort, South Africa
}

\begin{document}
\maketitle

\section{Introduction}

Since very-high-energy (VHE) gamma-ray sources emit characteristic photon fluxes of $\lesssim 10^{-12}$ erg cm$^{-2}$ s$^{-1}$, large collection areas $>100$ m$^{2}$ are required to detect them. Otherwise, the observations are statistically insignificant due to too low rates of observed photons.
Presently, it is not feasible for gamma-ray instrumentation with a physical size exceeding $\sim$1 m$^{2}$ to be sent into space, severely limiting the performance of direct observations in this energy range. The Imaging Atmospheric Cherenkov Telescopes (IACTs) take advantage of Cherenkov radiation visible in the optical range from 300 to 600 nm and allow indirect, ground-based observations of VHE radiation. The technique relies on the detection of short Cherenkov light flashes lasting 3--4 ns caused by the air showers generated during collisions of ultrarelativistic gamma-ray photons or hadrons with nuclei in the upper layer of the Earth's atmosphere. The ground-based detectors register information, leading to a precise identification of the primary particles, i.e. to measure their initial energy, establish the direction of emission, and determine the nature of the source of radiation.\\
The Cherenkov telescope array\footnote{\url{https://www.cta-observatory.org/}} (CTA) \citep{Acharya2013} is an emerging, ground-based, IACT observatory located in two hemispheres, namely CTA-North on the Canary island of La Palma (Spain) and CTA-South in the Atacama Desert in Chile. The observatory will consist of three types of telescopes, namely large- (LST), medium- (MST), and small- (SST) sized telescopes, having mirror diametres of 23 m, 11.5 m, and 4.3 m, and the fields of view of 4.5\dg, 7\dg, and 8\dg, respectively. In this manner, the CTA observatory tends to cover both hemispheres in a wide range of photon energy between 20~GeV and 300~TeV, improving the sensitivity level at the reference energy of 1 TeV by an order of magnitude in comparison to currently operating Cherenkov observatories. The key scientific goals of the CTA observatory include understanding the origin of VHE emission from astrophysical sources, the existence and the role of relativistic cosmic particles in the Universe, searching for dark matter, and many other topics \citep{CTAC2019}.\\
The first operating CTA telescope is the LST-1 located in the CTA-North site at the Roque de Los Muchachos Observatory on La Palma \citep{Cortina2019}. The sub-array scheme assumes the construction of four such telescopes at the central part of the northern array that provides the dominant contribution to the full system sensitivity in the range of 20 to 150 GeV. The inauguration of the LST-1 was held in 2018 and the telescope is currently in the science-engineering phase. The LSTs have a wide scope of observations, including Galactic transients, active galactic nuclei, gamma-ray bursts, and other astrophysical sources. It is worth mentioning that a fraction of observations will be affected by clouds, which cannot be avoided. However, the gathered data can be improved in the analysis process. \\
We propose a geometrical model correction of an image, which aims to improve the gamma/hadron separation and shower direction reconstruction. The proposed method is enhanced with a bias fit which is independent of energy and impact.

\section{Simulations}

Based on the physical parameters describing individual particle cascades, such as shape and components, that are later on reflected in their Cherenkov light images, hadronic and electromagnetic showers can be effectively distinguished from each other. The electromagnetic air showers are more regular and homogeneous, while the hadronic ones are much wider, often divided into supplementary sub-cascades. The Monte Carlo (MC) simulations showed that the individual hadronic showers also differ from each other much more than in the case of the electromagnetic cascades, which do not change much from one particle distribution to another \citep{Hillas1996}. The gamma-ray-initiated showers are only a small fraction of the whole population of particle cascades. Instead, most of the cascades are initiated by hadrons which are later considered background events. Moreover, the secondary particle showers caused by hadrons generate much more muons than the electromagnetic cascades, thus they can be used as a distinguishing indicator between showers.

\begin{wrapfigure}{l}{0.5\textwidth}
  \begin{center}
      \includegraphics[width=0.55\textwidth]{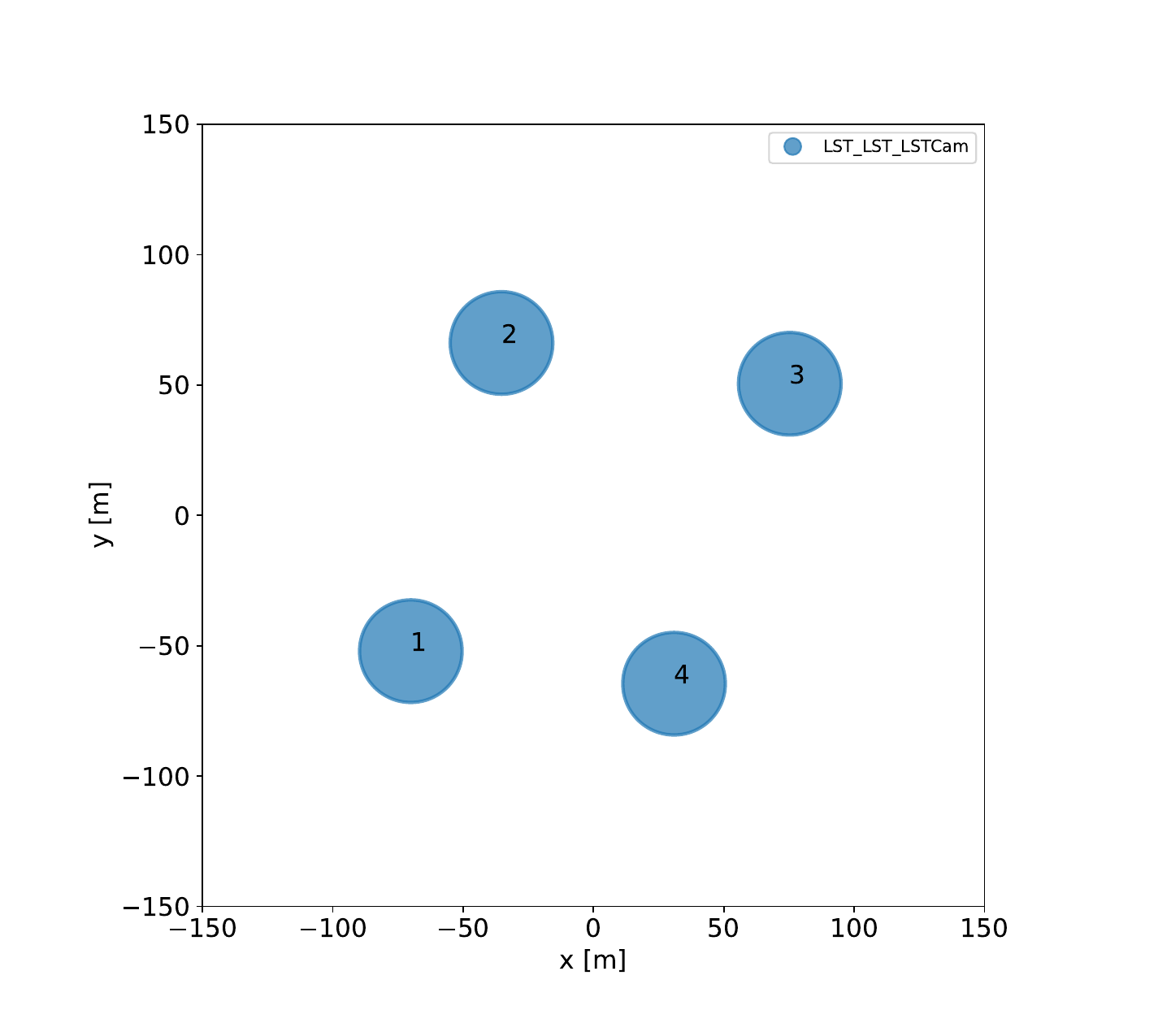}
  \end{center}
  \caption{The layout of four LSTs used in the MC simulations.} 
  \label{4LSTs}
\end{wrapfigure}

We have used the MC tool CORSIKA version 7.6 \cite{Heck:1998,corsika} to simulate the development of extensive air showers induced by gamma rays arriving from 20$\degree$ in zenith and 180$\degree$ in azimuth (counted clockwise from the geographic North). An additional package called    IACT/ATMO \cite{corsika} was used to simulate the emission of Cherenkov light in the wavelength range from 240 nm to 700 nm.  Gamma-ray-induced air showers, treated with the EGS4 \cite{osti_6137659} interaction model, were simulated in the center of the camera. The background protons, treated with hadronic models UrQMD \cite{Bass_1998} and QGSJET-II-04 \cite{Ostapchenko_2006} for, respectively, low-energy and high-energy interactions, were simulated arriving from a cone with a half-opening angle of 10$\degree$ around the aforementioned zenith and azimuth angles. In the second part of the MC production, another tool called \texttt{sim\_telarray} (Prod3b settings) is used \cite{corsika}. In this stage, the atmospheric impact on the propagation of Cherenkov photons and the response of the Cherenkov detector were simulated. For this purpose, we have simulated modified atmospheric profiles based on the U.S. 1976 Standard Atmosphere \cite{US1976} using the MODerate resolution atmospheric TRANsmission (MODTRAN) code version 5.2.2 \cite{stotts2019atmospheric}. Two atmospheric models were simulated: the clear atmosphere case containing no clouds or aerosol (total transmission is 100\%, referred to as $T = 1$), and the second model which includes a 1 km thick cloud with a base at 5 km a.g.l. (ground level is at 2147 m a.s.l.) and total transmission of 60\% ($T = 0.6$), which constitute a typical parameters of clouds on La Palma. The simulated layout consisting of 4 LSTs is shown in Figure~\ref{4LSTs}.  

\section{Geometrical model and data analysis}

To correct the data for the effect of the presence of clouds, we developed a simple geometrical model. 
Thanks to this we can apply a correction already at the image level. 
First, using uncorrected images, we perform standard image parametrization and stereoscopic reconstruction using the axis-crossing method. 
Comparing to the clear atmosphere case the parts of the image produce high in the atmosphere will be partially attenuated. Therefore the image will be less bright, and also the longitudinal distribution of the light will change.
Nevertheless, the direction of the main axis of the image should not be affected by the cloud  more than by the statistical fluctuations of light in individual pixels (that will be increased due to dimmer image). 
Therefore, such simple geometrical reconstruction would be more robust against the cloud effect to obtain the preliminary stereoscopic parameters. 

To improve the geometrical model, we have performed dedicated CORSIKA simulations of vertical gamma rays observed at different energies and impact parameters. 
We have applied a simple atmospheric absorption and computed the average offset angle of the Cherenkov photons from different ranges of emission height. 
To correct a minor bias (likely related to the latitudinal development of the shower), we introduce a phenomenological correction factor:
\begin{equation}
    \xi'= \xi\left(0.877 + 0.015 \frac{H + H_0}{7\,\mathrm{km}}\right), \label{eq1}
\end{equation}
where $H_0=2.2$~km is the height of the telescopes a.s.l., $\xi = \arctan(\cos(\theta) I/H)$ is the original offset angle corresponding to height $H$ (measured above the ground level), $\xi'$ is the corrected offset angle,  $\theta$ is the zenith angle of the observations, and $I$ is the preliminary reconstruction of the impact parameter.
Note that while the bias correction factor (the parentheses part of Eq.~\ref{eq1}) was optimized for low zenith angle observations, and thus might require modifications for observations at higher zenith distance angle, it does not depend on the energy of the primary gamma ray, nor the impact parameter. 

Next, we use the preliminary estimation of the arrival direction and impact to map each point on the camera to a specific height.
Namely, we split the camera in stripes perpendicular to the line connecting the preliminary source position and the center of gravity of the image (which represents the longitudinal development line). 
For each such stripe, we assign a single emission height: computed at the assumption that the Cherenkov photons are emitted from the axis of the shower.
We then correct the reconstructed signal in each pixel by an inverse of the transmission from the corresponding emission height. 
In the simple case of a thin cloud at height $H$ this is equivalent to computing the maximum offset angle from Eq.~\ref{eq1} corresponding to the cloud height ($H=H_{cl}$) and correcting only the pixels with a smaller offset angle by the inverse of the transmission.
The corrected image can be then parametrized again using the Hillas ellipse, and further stages of the analysis, i.e. the final reconstruction of the arrival direction, energy estimation, and the gamma/hadron event classification using the random forest (RF) method, can be performed. 

The aforementioned gamma- and proton-initiated simulations of observations gathered through the cloud at $H_{cl}$ = 5 km a.g.l with the transparency coefficient of $T$ = 0.6 and cloudless data were treated in the same way and analyzed in the \textsc{ctapipe}\footnote{\url{https://github.com/cta-observatory/ctapipe}} version 0.12.0 and \textsc{lstchain}\footnote{\url{https://github.com/cta-observatory/cta-lstchain}} version 0.9.2 frameworks. The reduction of raw data directly from the image server to the first data level was performed in a standard approach, using modified \textit{r0\_to\_dl1}\footnote{\url{https://github.com/cta-observatory/cta-lstchain/blob/master/lstchain/reco/r0_to_dl1.py}} script, where the Hillas parameters from the image, the tentative stereo reconstruction parameters, impact, and the source position were calculated. Subsequently, we corrected images with the geometrical model and then recalculated the Hillas as well as stereo parameters. Eventually, we estimated the second level of parameters, including gamma/hadron separation parameter (\textit{gammaness}) and reconstruction of the parent particle parameters (energy and direction), using the RF method. The RF models were trained based on 849,509 events in total, using the 50\% to 50\% ratio of the training to testing sub-sample of cloudless data and, finally, applied to the MC simulations with the cloud.

\section{Results}

\begin{figure}[!ht]
\centering
\includegraphics[width=1.\textwidth]{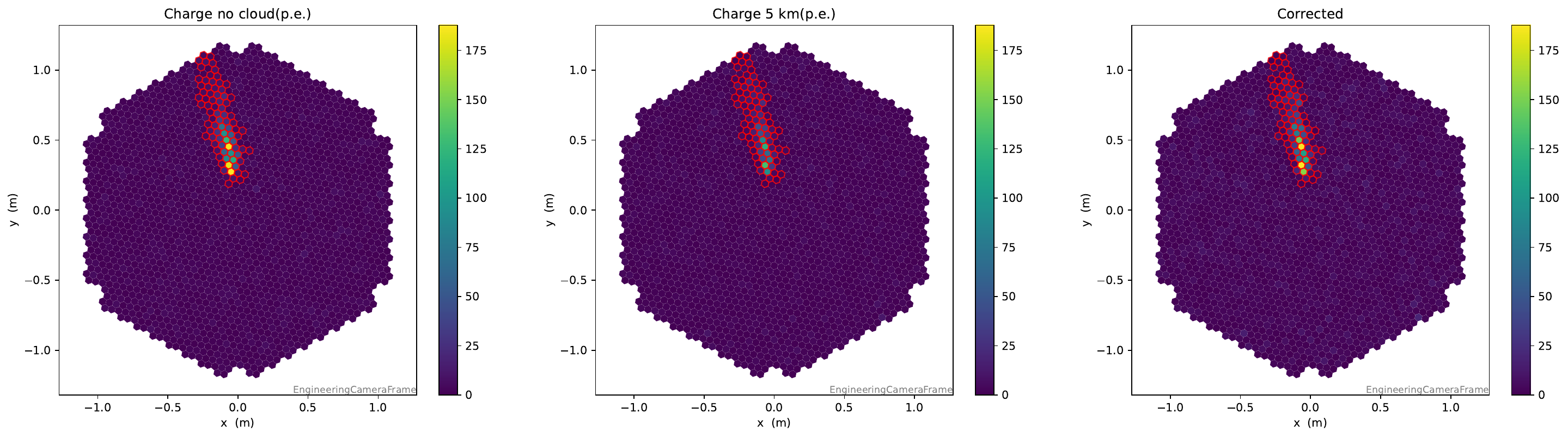}
\includegraphics[width=1.\textwidth]{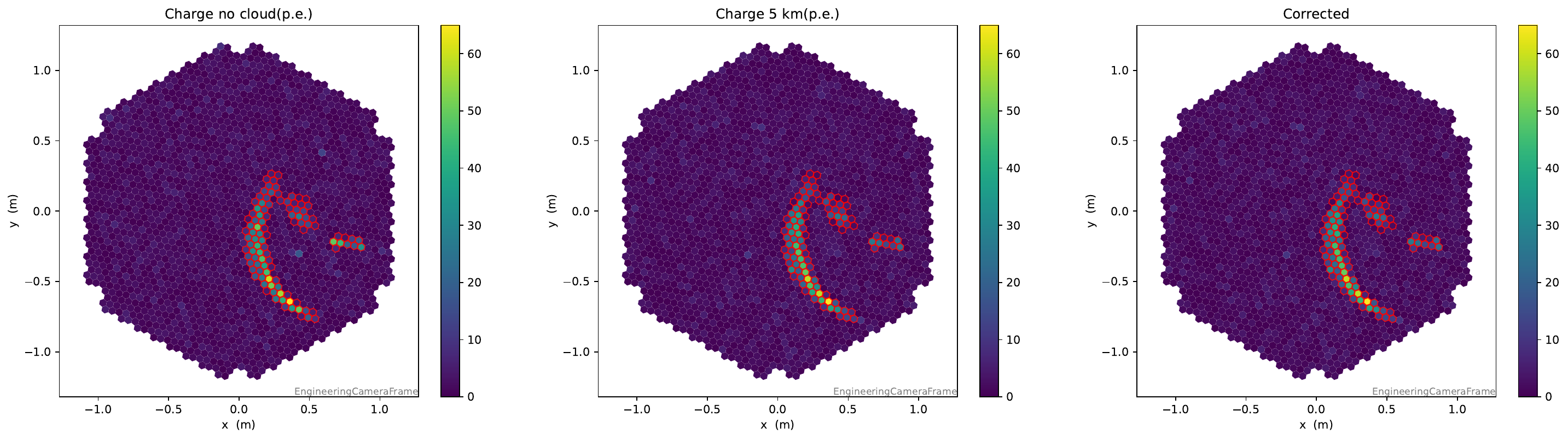}
\caption{LST image of electromagnetic shower (upper panel) and hadronic shower (bottom panel). \textit{Left-hand side:} LST image without a cloud, \textit{middle part:} image with a cloud at 5 km a.g.l. with $T = 0.6$, and \textit{right-hand side:} image with the cloud after correction.}
\label{gammas_protons}
\end{figure}

As the result of our analysis approach, we obtained a few sets of images and parameters. Figure~\ref{gammas_protons} stands as an example of the LST camera images of gamma- (top panel) and proton-initiated (bottom panel) MC simulations of the cloudless sky (left panel), affected by the cloud (middle panel), and in the presence of the same cloud with the geometrical model applied (right panel). This gives us an indication of how the image on the camera changes depending on the data used and a very preliminary glimpse of the cloud correction effect.

\begin{figure}[!ht]
\centering
\includegraphics[width=1.\textwidth]{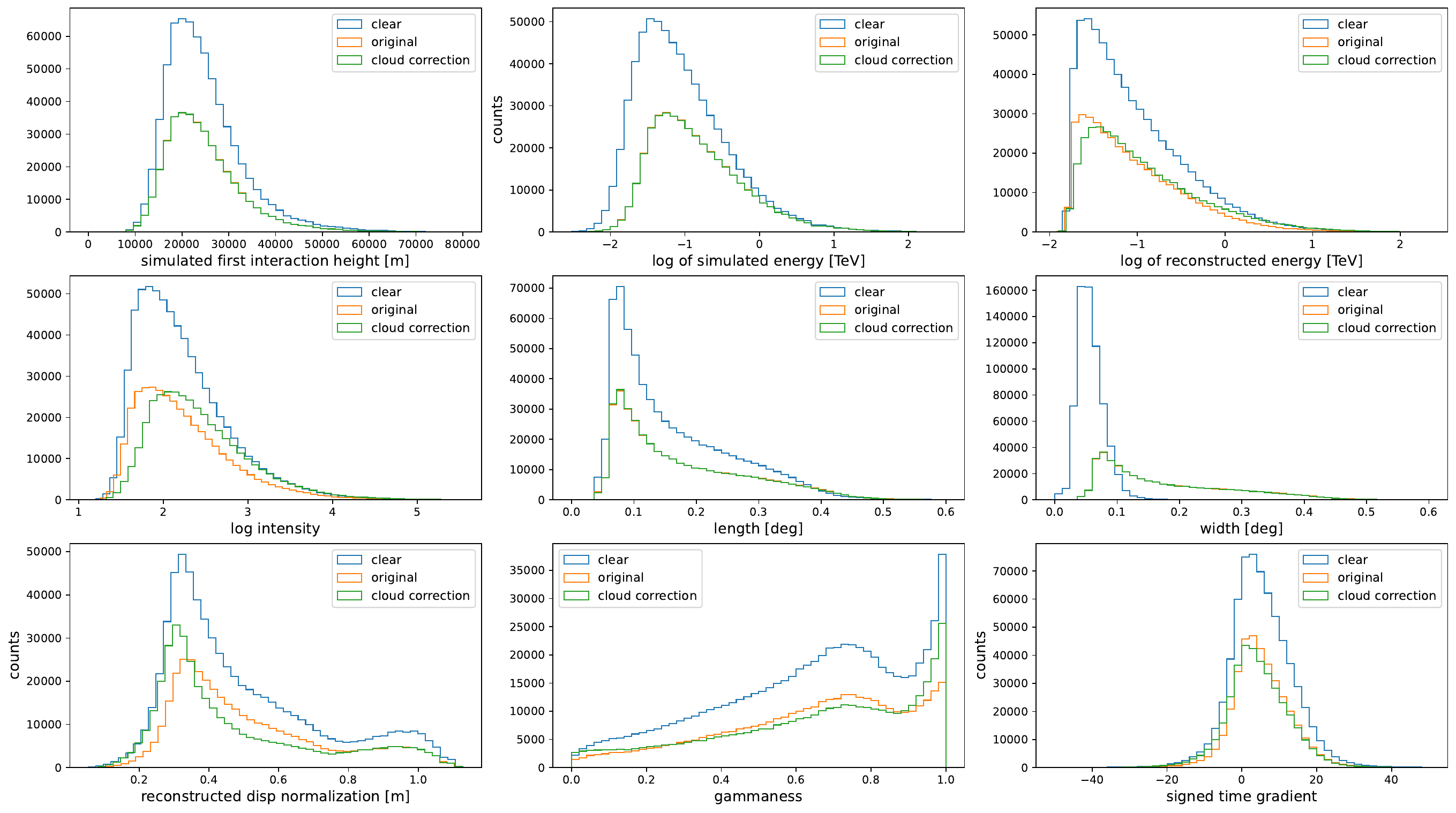}
\caption{Distributions of parameters of the MC simulated gammas for $T = 1$ (clear), $T = 0.6$ for the cloud at 5 km a.g.l. (original), and for the same cloud with correction model applied (cloud correction).}
\label{dl2_params}
\end{figure}

\begin{figure}[!ht]
\centering
\includegraphics[width=1.\textwidth]{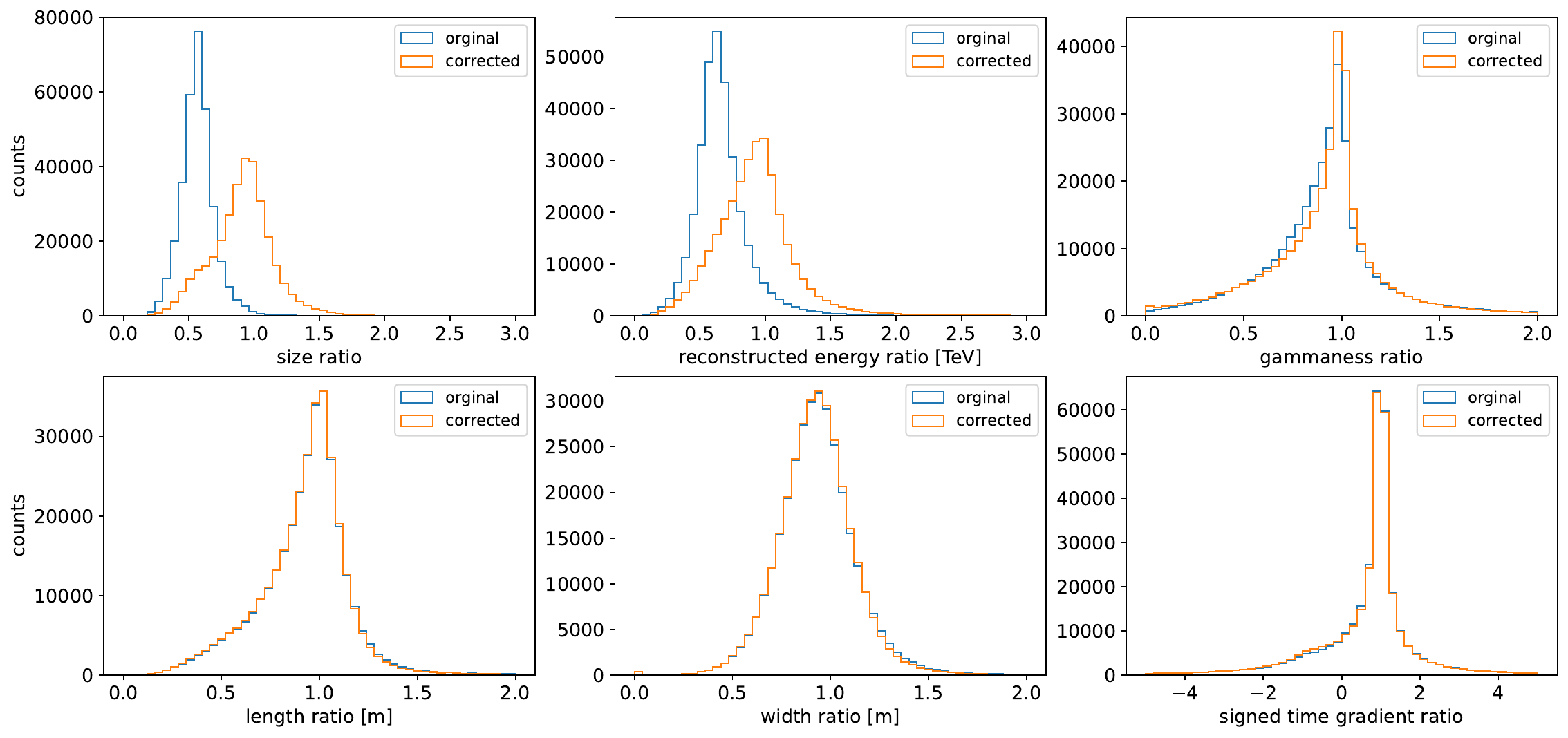}
\caption{Ratios of original and corrected parameters of the MC simulated shower for $T = 0.6$ to $T = 1$ for the cloud at 5 km a.g.l.}
\label{dl2_ratios}
\end{figure}

Recalculating the Hillas parameters from the corrected images, we should be able to improve the size (total intensity), shape (width and length), the height of the first interaction of the particle, and reconstructed energy of each event with the respect to the cloudless MC simulations. Figure~\ref{dl2_params} displays histograms of the MC simulations of gamma-initiated shower reconstructed parameters for all studied cases. It is visible that the distributions of DL\,2 parameters derived based on the simulations affected by the cloud vary from the corrected ones, except for the simulated energy and the first interaction height. A similar effect is also visible in ratios of the same parameters to the parameters estimated from the cloudless simulations (see Figure~\ref{dl2_ratios}). 
In the case of size and reconstructed energy the ratio show a clear bias (comparable to the transmission of the cloud), that gets corrected with the proposed method.
Instead, the length and with parameters do not show a bias and only a spread when compared to the clear atmosphere case. 
This might be related to the fact that the cloud assumed in those preliminary simulations is relatively low, hence affecting most of the image in the same way.

\section{Summary}

We have generated and analyzed a few sets of the MC simulations of gamma-ray- and proton-initiated showers observed by an array of four LSTs under two different atmospheric conditions, i.e., without and with the 60\% opaque cloud based at 5 km a.g.l. In our approach, we developed a simple geometrical model to correct for the cloud presence during the data gathering, which provides promising initial improvements in the reconstruction of the shower parameters. However, we considered a simple form of a uniform in shape and over time cloud, which could be treated as the first approximation to the problem. In the next steps, we intend to take into account more complex cases, including different cloud transparency and  height, as well as adding the wobble mode in the MC simulations to get more realistic data. The results we obtained are comparable to the results obtained by the method used by \cite{Dorner2009}. The presented method corrects the images directly without the need for time-consuming and resource-expensive dedicated MC simulations.

\end{document}